\documentclass[useAMS,usenatbib]{mn2e}

\usepackage{multicol}
\usepackage{graphicx}
\usepackage{color}
\usepackage{graphicx}
\usepackage{amssymb}
\usepackage{amsmath}
\usepackage{rotating}
\usepackage{longtable}
\usepackage{lscape}

\title[The globular cluster 2MASS-GC\,03]{Near-infrared photometry and spectroscopy of the  low Galactic latitude globular cluster 2MASS-GC\,03}
\author[Carballo-Bello et al.]
{Julio A. Carballo-Bello$^{1,2,8}$
\thanks{E-mail: julio.carballo@uv.cl},
S. Ram\'irez Alegr\'ia$^{1,2}$, J. Borissova$^{1,2}$, L. C. Smith$^{3}$,  
\newauthor
R. Kurtev$^{1,2}$, P. W. Lucas$^{3}$, Ch. Moni Bidin$^{4}$, J. Alonso-Garc\'ia$^{5,2}$, { D. Minniti$^{6,2,7}$}, 
\newauthor
{T. Palma$^{2,6}$, I. D\'ek\'any$^{8,2}$},  N. Medina$^{2,1}$,
M. Moyano$^{4}$, V. Villanueva$^{2,1}$ and M. A. Kuhn$^{1,2}$
\\
\\
$^{1}$Instituto de F\'isica y Astronom\'ia, Facultad de Ciencias, Universidad de Valpara\'iso,\\ Av. Gran Breta\~na 1111, Playa Ancha, Casilla 5030, Valapara\'iso, Chile\\
$^{2}$Millenium Institute of Astrophysics, Av. Vicu\~na Mackenna 4860, 7820436, Macul, Santiago, Chile\\
$^{3}$Centre for Astrophysics Research, Science and Technology Research Institute, University of Hertfordshire, Hatfield AL10 9AB, UK\\
$^{4}$Instituto de Astronom\'ia, Universidad Cat\'olica del Norte, Av. Angamos 0610, Antofagasta, Chile\\
$^{5}$Unidad de Astronom\'ia, Facultad de Cs. B\'asicas, Universidad de Antofagasta, Av. U. de Antofagasta 02800,
Antofagasta, Chile\\
$^{6}$Departamento de F\'isica, Facultad de Ciencias Exactas, Universidad Andr\'es Bello, Av. Fern\'andez Concha 700, Las Condes, Santiago, Chile\\
$^{7}$Vatican Observatory, V00120 Vatican City State, Italy\\
$^{8}$Instituto de Astrof\'isica, Facultad de F\'isica, Pontificia Universidad Cat\'olica de Chile (IA-PUC), Casilla 306, Santiago 22
}

\usepackage{natbib}
\newcommand{\figref}{Figure~\ref}
\newcommand{\tabref}{Table~\ref}

\bibpunct{(}{)}{,}{a}{,}{;}
\begin{document}

\date{Accepted ??}

\pagerange{\pageref{firstpage}--\pageref{lastpage}} \pubyear{2002}

\maketitle

\begin{abstract}
We present deep near-infrared photometry and spectroscopy of the globular cluster 2MASS-GC\,03 projected in the Galactic disk using MMIRS on the Clay telescope (Las Campanas Observatory) and VISTA Variables in the Via Lactea survey (VVV) data. Most probable cluster member candidates were identified from near-infrared photometry. Out of ten candidates that were followed-up spectroscopically, five have properties of cluster members, from which we calculate $<$[Fe/H]$> = -0.9 \pm 0.2$ and a radial velocity of $<v_{\rm r}> = -78 \pm 12$\,km/s. A distance of 10.8\,kpc is estimated from 3 likely RR\,Lyrae members. Given that the cluster is currently at a distance of 4.2\,kpc from the Galactic center, the cluster's long survival time of an estimated $11.3\pm 1.2$\,Gyr strengthens the case for its globular-cluster nature. The cluster has a hint of elongation in the direction of the Galactic center.
\end{abstract}


\section{Introduction}

Galactic globular clusters (GCs) have played a role in diverse areas of study, from the evolution of stellar populations to the hierarchical formation of galaxies. The current population of known GCs in the Milky Way is composed of $\sim 160$ of these systems \citep{Harris2010}, although it has long been suggested \citep[e.g.][]{Ashman1992} that a fraction of the total number of Galactic GCs might be hidden in lines of sight with high extinction (i.e. disk, bulge) or have intrinsic properties that make them difficult to detect. The search for new Galactic GCs has only returned a small number of the expected missing GCs. Some of these newly identified GCs have properties mid-way between a GC and an ultra-faint dwarf galaxy \citep{Carraro2005,Willman2005,Koposov2007,Longmore2011,Munoz2012,Balbinot2013,Laevens2014,Kim2015}. Most of these new detections were possible thanks to the arrival of wide-sky photometric surveys such as Sloan Digital Sky Survey \citep[SDSS,][]{York2000}, the Two Micron All Sky Survey \citep[2MASS,][]{Skrutskie2006} and the VISTA Variables in the
Via Lactea (VVV) Public Survey \citep{Minniti2010,Saito2012}. The latter survey reported the discovery of VVV CL\,001
\citep{Minniti2011}, VVV CL\,002 \citep{MoniBidin2011} and several other GC candidates \citep{Borissova2014}. 

It is well known that the Galactic GC system contains a metal-rich subsystem \citep{Zinn1985} which is mainly composed of GCs associated with the Galactic bulge. Among the 13 GCs found with $|b| < 2^{\circ}$ (8$\%$ of the Galactic population), 11 lie projected on the Galactic bulge, while only two, 2MASS-GC\,03 and GLIMPSE\,01 \citep{Simpson2004,Kobulnicky2005,Davies2011}, are likely associated with the Milky Way disk. However, the nature of GLIMPSE\,01 remains unclear \citep[see][]{Davies2011}. In contrast, numerous young cluster candidates have been discovered in the Galactic plane using infrared photometry \citep[e.g.][]{Froebrich2007a,Borissova2011,Borissova2014,Solin2014}. Therefore, the scarce GCs at low Galactic latitudes represent a peculiar population of stellar systems in a region of the Milky Way dominated by the presence of young stellar clusters. 

  \begin{figure*}
     \begin{center}
      \includegraphics[scale=0.37]{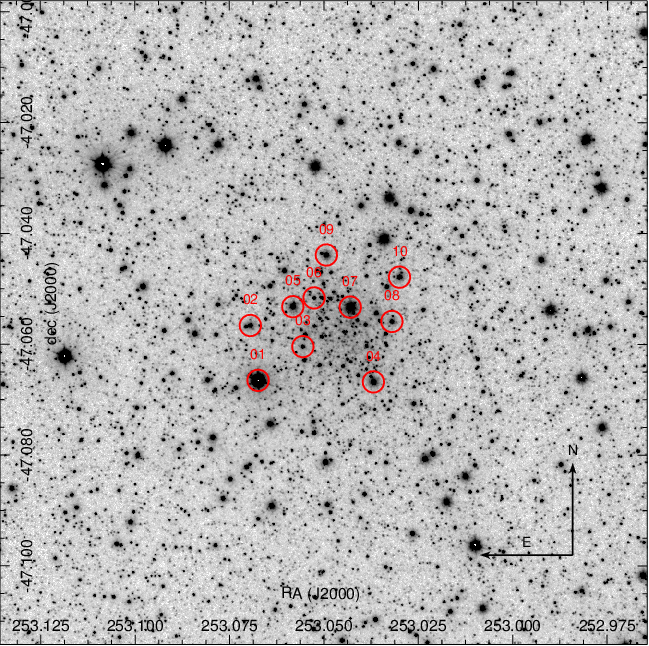}
      \includegraphics[scale=0.37]{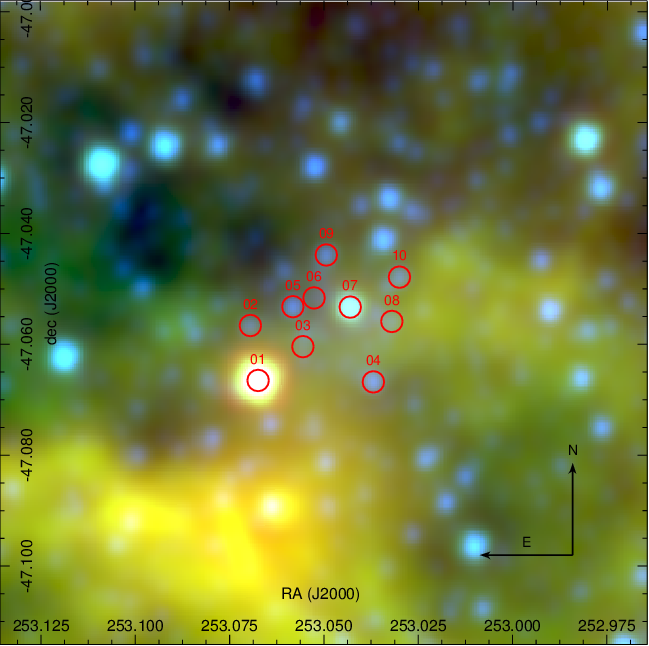}
      \caption[Imagenes]{VVV $K_S$ ({\it left}) and WISE false color ({\it right}) images for GC\,03. The red circles show the position of the stars observed spectroscopically. The false color image was constructed using the WISE filters {\it W}1 (blue), {\it W}2 (green) and {\it W}3 (red). The image sizes are 7\,arcmin $\times$ 7\,arcmin.}
\label{fcr1735_VVV_WISE}
     \end{center}
   \end{figure*}

\citet{Froebrich2007a} first identified 2MASS-GC\,03 as a local density enhancement in 2MASS photometry maps, listing it as \#1735 (or FSR\,1735) in their catalog. Follow-up $JHK_s$ photometry of stars associated with that overdensity \citep{Froebrich2007b} allowed the authors to suggest that FSR\,1735 is actually a GC at $d_{\odot} \sim 9$\,kpc, although they point out that a definitive determination of the object's nature would require follow-up observations. A systematic analysis of the cluster was performed by \cite{Kharchenko2013}, as a part of a global investigation on the Galactic star-cluster system. Using the 2\,MASS catalogs and PPMXL Catalogue of Positions and Proper Motions on the ICRS \citep{Roeser2010}, they estimated the metallicity, distance and age of FSR\,1735 as [Fe/H] = -0.6, $d_{\odot} \sim 8.8$\,kpc and $t = 9.9$\,Gyr, respectively. Because of its discovery in 2\,MASS data, FSR\,1735 has the SIMBAD denomination 2\,MASS-GC03 and it also corresponds to the candidate number 71 in the \cite{Solin2014} catalog, which contains new star cluster candidates discovered in the VVV field of view. In this paper, we will use the SIMBAD designation, namely 2\,MASS-GC\,03 or simply GC\,03. As it can be seen in Figure \ref{fcr1735_VVV_WISE}, we observe a clear overdensity of stars in the VVV $K_s$-image, but it is not associated with mid-infrared nebulosity in WISE images (Figure \ref{fcr1735_VVV_WISE}, right panel), confirming the Froebrich et al.\ conclusion of small amounts of dust and gas in that area. 

In this paper, we present deep MMIRS and VVV photometry and low-resolution near-infrared spectroscopy of several stars in order to confirm the globular nature of the cluster GC\,03 and derive accurate values for the fundamental parameters that describe that system.

\section{Observations and Data Reduction}

The analysis of GC\,03 \citep[RA=$16^h52^m10.6^s$, DEC=$-47^{\circ}03\arcmin29\arcsec$][]{Harris2010} includes near-infrared images and spectra, both obtained using the MMT and Magellan Infrared Spectrograph \citep[MMIRS,][]{mmirs12} at the Clay Telescope (Las Campanas Observatory, Chile). The instrument is a wide field near-infrared imager and spectrograph. We have complemented our data using the VVV $JHK_s$ photometry.

With the MMIRS, the cluster was observed through $J$ and $K_s$ filters, with a total exposure time per filter of 119 seconds and airmasses between 2.35 and 2.08. We used a dithering pattern with offsets in x and y-axis of 34 arcsec; these offsets in the images are used later to construct the calibration image for sky subtraction. We reduced the imaging data using {\sc iraf}\footnote{{\sc iraf} is distributed by the National Optical Astronomy Observatories (NOAO),  which is operated by the Association of Universities for Research in Astronomy, Inc. (AURA) under cooperative agreement with the U.S.A. National Science Foundation (NSF).} tasks (dark subtraction, flat fielding, and sky correction) and we used python package ALIPY\footnote{http://obswww.unige.ch/$\sim$tewes/alipy/} to align the individual reduced images before the final combination. 

The PSF photometry was independently derived for each of the images using DAOPHOT\,II/ALLSTAR \citep{Stetson1987}. The calibration of the photometric catalogs was performed using the $\sim$1000 stars in common with 2MASS, and we substitute 2MASS photometry for objects that are saturated or present other problems in the PSF photometry procedure. The magnitudes obtained from individual images were averaged for the final catalog. 
To estimate completeness, artificial stars were inserted randomly throughout the chip in the range $10 < K_s < 20$~mag, and these were extracted using the same photometric procedure. The total number of synthetic stars added was designed not to exceed 15$\%$ of the number of natural sources. This procedure was performed 50 times for each frame. The average magnitude limits of our catalogs, where the fraction of synthetic stars recovered is the 50\%, are 18.6~mag and 17.6~mag in $J$ and $K_s$, respectively.  

To select promising candidates for spectroscopy, we used PSF photometry from the VVV-SkZ pipeline \citep{Mauro2013}, an automated software based on ALLFRAME \citep{Stetson1987} and optimized for VISTA PSF photometry. We used $J$ and $K_s$ images from the VVV survey \citep{Minniti2010} and replaced the photometry from saturated sources (i.e. brighter than $K_s = 12$), with 2MASS photometry. Position and near-infrared colors and magnitudes were used to remove likely contaminants from the photometric catalog using the algorithm of \citet{bonatto10}, as described by \citet{Borissova2011}. The preliminary decontaminated CMDs allowed us to select bright ($6.56 < K_s < 11.82$) cluster member candidates for multi-object spectroscopy. We selected 10 stars to observe with a single mask with MMIRS, using the $HK$ grism, a slit width of 0.4\,arcsec, and a resolution $R\sim1400$. 

\begin{figure*}
     \begin{center}
      \includegraphics[scale=0.8]{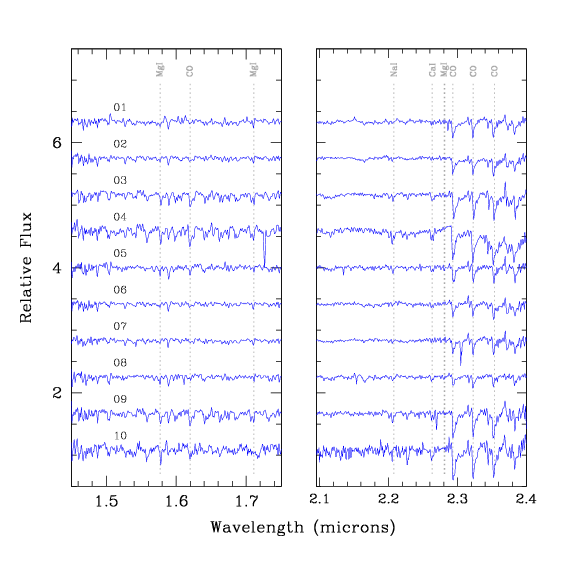}
      \caption[Imagenes]{Individual spectra in $H$ (left) and $K$ band (right) for the observed stars. We marked with grey vertical lines the spectral lines used for metallicity and radial velocity estimates.} 
\label{spectra}
     \end{center}
   \end{figure*}

Standard IRAF procedures were used to de-bias and flat-field the spectra. The 10 one-dimensional spectra were extracted, and the wavelength calibration was obtained using a third-order polynomial fit to $\sim$40 arc lines, resulting in a dispersion of $\sim 6.7$\AA\,pix$^{-1}$ and a rms scatter of 0.4\AA. We corrected the atmospheric absorption lines with MOLECFIT \citep{Kausch2015,Smette2015}, a software tool distributed by ESO which uses the information available in the spectra header to model the atmospheric and telescope conditions and the telluric lines during the observations. The resulting spectra are shown in \figref{spectra}.

\section{Results}

The cluster center was determined with an iterative procedure based on a VVV photometric catalog of stars brighter than $K_s=17$. We first calculated the barycenter of all the sources detected within 1\,arcmin from the \cite{Froebrich2007b} coordinates. The point thus found was adopted as new cluster center, and the calculation iterated. The procedure quickly converged to fluctuations smaller than one pixel ($0\farcs33$) around the values RA=$16^h52^m11.1^s$, DEC=$-47^{\circ}3\arcmin30\arcsec$, which represents our best estimate of the cluster center. This is only $7\farcs6$ offset from \cite{Froebrich2007b} coordinates.

\subsection{Decontaminated color-magnitude diagram}
\label{decontamination}

To determine the most probable cluster members we used the deeper MMIRS photometry.
As expected for low Galactic latitudes and poorly populated targets, the near-infrared point-source  catalogs were dominated by field stars. Therefore statistical decontamination of the photometric catalog is necessary to provide a cleaner sample of the GC\,03 stellar population that will be used for further analysis. The algorithm we use (proposed by \cite{Bonatto2007}), is based on a multivariate search for over-densities in the $JHK_s$ catalogs. Although the cluster membership cannot be determined absolutely, the algorithm is rather effective at removing the majority of field stars. As described in more detail in \cite{Bonatto2007} paper, the algorithm divides the ($K_s, J - K_s$) color-magnitude space into a grid of cells with a cell size provided as input. It estimates the expected number density of cluster stars for each of the cells by subtracting the field stars number density in the same ($K_s, J - K_s$) ranges and, summing over all cells, it obtains a total number of candidate stars $N_{\rm mem}$. The procedure is repeated with the grid shifted in steps of $1/3$ cell width in each dimension, yielding 9 different configurations, with nine different  $N_{\rm mem}$ values. 
 Those stars that {\it survive} to this process more than  $<N_{\rm mem}>$ times are considered likely members. For this work, we slightly modified the procedure by including a family of cell sizes both for $K_s$ and $J - K_s$ in the range $0.2 < \delta K_s < 0.6$ and $0.1 < \delta(J - K_s) < 0.3$. 
 We generated a list of tentative cluster members derived from 15 iterations, based on the number of times that a star was present in the results. As a final step, we ranked the stars depending on that survival frequency and assigned a probability, $P_{\rm memb}$, to each star based on their frequency of survival. Thus, it is possible to select stars using a different $P_{\rm memb}$ threshold for more or less reliable subsamples. The procedure was applied within a radius of 1.5~arcmin of the cluster center, which contains, from visual analysis, the greatest density of stars. 

  \begin{figure*}
     \begin{center}
      \includegraphics[scale=0.47]{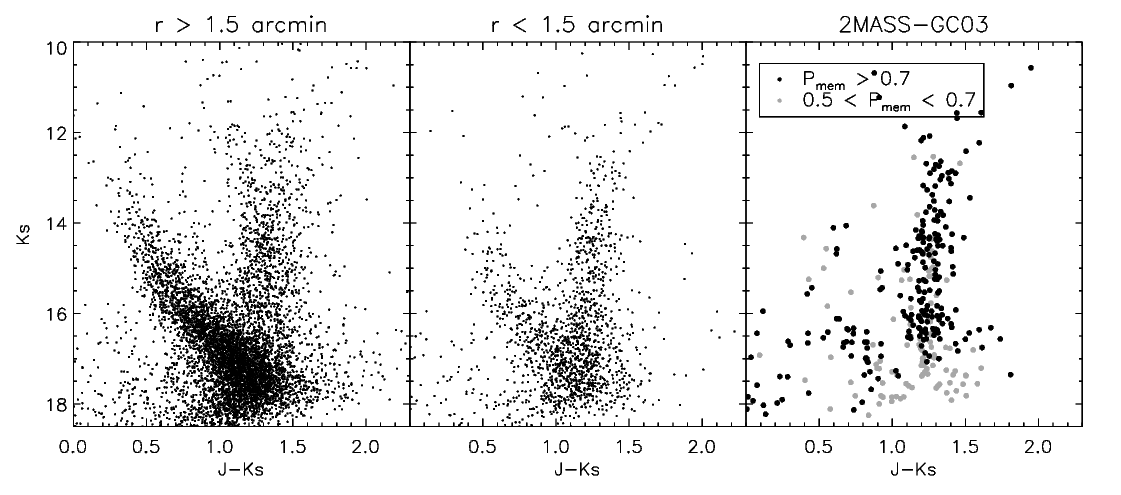}
      \caption[CMDs]{\small CMDs corresponding to stars beyond and within 1.5\,arcmin of the cluster center (left and middle panels respectively) and the resulting decontaminated CMD (right panel). The latter panel displays stars with higher probability of membership ($P_{\rm mem} > 0.7$) as black dots while stars with lower membership probability ($ 0.5 >P_{\rm mem} > 0.7$) are shown in grey.}
\label{cmds}
     \end{center}
   \end{figure*}

\figref{cmds} shows the importance of the decontamination process to obtain useful CMDs. From the initial 1649 sources within 1.5\,arcmin from the center of GC\,03 (middle panel), our algorithm found 348 candidate stars for any $P_{\rm mem}$ value. Focusing on those stars with the highest membership probabilities, $P_{\rm mem} > 0.5$, the sample of member stars is reduced to 315 sources. 

\begin{figure*}
     \begin{center}
     \includegraphics[scale=0.7]{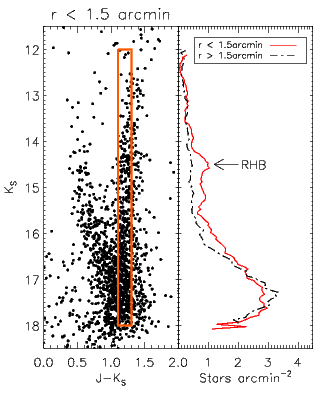}
      \caption[RedClump]{{\small {\it Left:} CMD corresponding to all stars contained in the area with $r < 1.5$\,arcmin from the center of GC\,03. The overplotted orange box indicates the region of the CMD used to derive the luminosity functions. {\it Right:} luminosity function for  all stars with $r < 1.5$\,arcmin (red) and field stars beyond 1.5\,arcmin from the cluster center (dotted black line). The position of RHB is indicated with an arrow.}} 
\label{redclump}
     \end{center}
   \end{figure*}

As shown in \figref{cmds}, the decontaminated CMD shows a well defined Red-Giant Branch (RGB) in the range $12 < K_s < 16.5$ mag. A  conspicuous overdensity is found at $K_s \sim 14.4$ mag that could be composed of Red Horizontal Branch stars (HB). However, the detection of Milky Way Red Clump (RC) stars in our field is also expected given that they are the dominant population among the Galactic giant stars. In fact, they have been long used to trace the Galactic structure because of their narrow luminosity function \citep[e.g.][]{Cabrera-Lavers2008,McWilliam2010,Gonzalez-Fernandez2014}. To confirm the association of that overdensity of stars with the cluster, we constructed area-normalized luminosity functions by counting stars in the range $1 < J-K_s < 1.5$ and $12 < K_s < 18$ with a bin size of $\delta K_{S} = 0.2$ for all the sources within 1.5\,arcmin from the cluster center and for those stars beyond 1.5\,arcmin. The results (see \figref{redclump}) show a discrepancy between the distributions for the field and cluster areas and a peak is found at $K_s \sim 14.4$ mag. The lack of a significant component of field stars in that position confirms that only a stellar population in the $r < 1.5$\,arcmin region, lying in an extremely narrow distance range, is able to generate such a  well-defined feature in the CMD. Thus, we conclude that the overdensity detected in $K_s \sim 14.4 \pm 0.2$ mag corresponds to the RHB of GC\,03, also detected by Froebrich et al., but 0.4 mag fainter in $K_s$. The distributions shown in \figref{redclump} suggests an additional second peak at $K_s \sim 15.1$. This was suggested by \cite{Froebrich2007b}, although they reported that its presence depended on the magnitude bin used to construct the luminosity function. The profile of the peak varies slightly with different bin sizes but the double peak morphology is always observed. As for the RC, the absence of a significant variation of stellar counts in the field around GC\,03, suggests that this hypothetical second overdensity is associated with the cluster. We propose that these stars are crossing the so-called Red Giant Branch Bump (RGBB), a transitory variation in the magnitude of RGB stars produced by a discontinuity in the chemical abundance profile, which is detected in a significant number of Galactic GCs \citep[e.g.][and references therein]{Valenti2007,Nataf2013}. However, the additional investigation of RGBBs is beyond the scope of our work with the data available. The peak around $K_s = 12.6$ mag, which appear in  the Froebrich et al.\ CMD and was interpreted as AGB bump, is not found. 

\subsection{Metallicity and radial velocity}

To calculate the [Fe/H] values of the spectroscopically measured stars, we used the metallicity calibration by \citet{Frogel2001}, based on the Equivalent Width (EW) of the Na\,I 2.21 $\mu m$ and Ca\,I 2.26 $\mu m$ lines and the ${}^{12}$CO\,(2,0) band. The calibration is particularly useful for moderate-resolution $K$-band spectra and was completed using spectra from 15 GCs. We measured the EW of our targets on the continuum normalised spectra, with the {\sc iraf} task {\it splot}. In general, we fit line profiles using a Gaussian with a linear background, and in more complicated profiles we used the deblending function or Voigt profile fitting. The EW uncertainties were estimated taking into account the S/N of the spectra, the peak to continuum ratio of the line and the error from the telluric star subtraction (which was estimated to be $\sim$10\% in the worst cases). The uncertainties of the \citet{Frogel2001} calibration are assumed to be 0.2 dex.   

We measured the radial velocities (RVs) of the targets with the {\sc iraf} task {\it rvidlines}, which fits spectral lines to determine  the wavelength shift with respect to specified rest wavelengths. In this procedure, we used the Mg\,I (at 1.5770, 1.7109 and 2.2813 $\mu m$), Na\,I (at 2.2062 and 2.2090 $\mu m$), Ca\,I (at 2.2631 and 2.2657 $\mu m$) lines and the ${}^{12}$CO\,(6,3), (2,0) and (3,1) bands, available in the spectra (typically 7--10 lines per spectrum). To calculate uncertainties, we added in quadrature the line-measurements errors from {\it rvidlines} and the errors from the wavelength calibration of each spectrum. Table~1 gives the metallicity and radial velocities for the stars with spectroscopic follow-up.

\begin{table*}
\centering
\begin{tabular}{lllrrrrrr}
\hline
ID &        RA &         DEC &       $K_s$ &          $J-K_s$ &           [Fe/H]&            RV  &           $\mu_{\alpha\cos\delta}$  &  $\mu_{\delta}$\\
   &  [J2000] & [J2000] &  & & & [km\,s$^{-1}]$ & [mas\,y$^{-1}$] & [mas\,y$^{-1}$]\\
\hline
  01*  & 16:52:16.2 & -47:03:59.5 & 6.56  $\pm$ 0.03 & 2.40 $\pm$ 0.04 & $-$0.95 $\pm$ 0.18 & $-$95  $\pm$ 11  & $---$                & $---$              \\
  02*  & 16:52:16.7 & -47:03:23.9 & 10.32 $\pm$ 0.04 & 2.14 $\pm$ 0.06 & $-$0.93 $\pm$ 0.20 & $-$83  $\pm$ 12  & $+$0.06 $\pm$ 0.88   & $-$2.35 $\pm$ 0.89   \\
  03   & 16:52:13.4 & -47:03:37.6 & 11.34 $\pm$ 0.03 & 1.18 $\pm$ 0.04 & $-$0.18 $\pm$ 0.33 & $+$112 $\pm$ 14  & $+$5.01 $\pm$ 0.57   & $+$0.95 $\pm$ 0.58    \\
  04   & 16:52:08.9 & -47:04:00.5 & 9.54  $\pm$ 0.02 & 1.9  $\pm$ 0.2  & $+$0.74 $\pm$ 0.49 & $-$60  $\pm$ 10  & $---$                & $---$              \\
  05   & 16:52:14.0 & -47:03:11.7 & 9.60  $\pm$ 0.03 & 1.98 $\pm$ 0.04 & $-$0.57 $\pm$ 0.26 &   0    $\pm$ 9   & $---$                & $---$              \\
  06*  & 16:52:12.6 & -47:03:06.0 & 11.84 $\pm$ 0.07 & 2.53 $\pm$ 0.05 & $-$0.76 $\pm$ 0.22 & $-$75  $\pm$ 10  & $+$1.42 $\pm$ 0.58   & $-$1.60 $\pm$ 0.59   \\
  07*  & 16:52:10.3 & -47:03:11.9 & 8.21  $\pm$ 0.04 & 2.53 $\pm$ 0.05 & $-$0.97 $\pm$ 0.20 & $-$72  $\pm$ 9   & $---$                & $---$              \\
  08*  & 16:52:07.7 & -47:03:21.2 & 10.63 $\pm$ 0.03 & 2.31 $\pm$ 0.04 & $-$0.82 $\pm$ 0.22 & $-$63  $\pm$ 11  & $-$1.95 $\pm$ 0.61   & $-$4.33 $\pm$ 0.60  \\
  09   & 16:52:11.8 & -47:02:38.1 & 9.85  $\pm$ 0.03 & 2.02 $\pm$ 0.04 & $-$0.04 $\pm$ 0.34 &   0    $\pm$ 12  & $+$13.77 $\pm$ 1.75  & $+$17.24 $\pm$ 1.75  \\
  10   & 16:52:07.2 & -47:02:52.5 & 9.99  $\pm$ 0.02 & 1.98 $\pm$ 0.03 & $+$0.22 $\pm$ 0.40 & $+$14  $\pm$ 20  & $-$1.35 $\pm$ 0.99   & $+$4.79 $\pm $1.01   \\
\end{tabular}
\label{metal_table}
\caption[Map]{$K_s$, $K_s - J$, metallicity, radial velocity and relative proper motions derived for the stars with spectroscopic follow-up.
Asterisks indicate likely cluster members. The proper motions are relative.}
\end{table*}

\subsection{Proper motions of GC\,03}
\begin{figure*}
 \begin{center}
	\vspace{0.5cm}
      \includegraphics[scale=0.45]{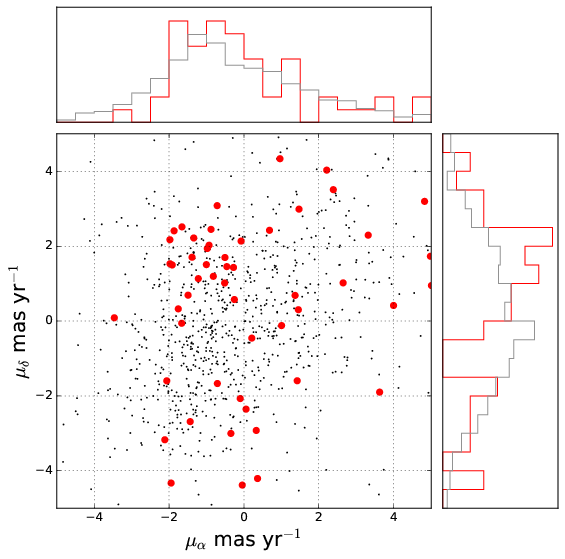}
      \includegraphics[scale=0.45]{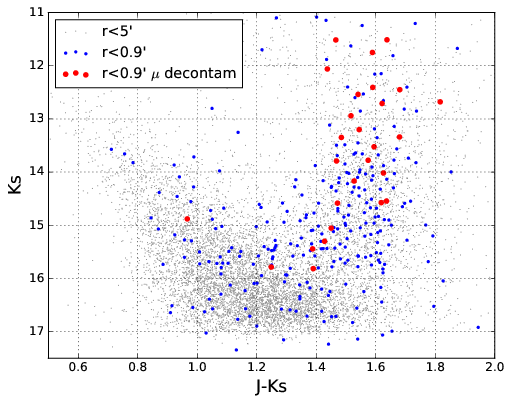}
     \caption[pm]{{\small {Left panel: Pipeline proper motions of $K_s < $14 mag sources within 5 arcmin (grey points), and 0.9 arcmin (red points) of the globular cluster GC03. Notice the concentration of red points in PM space. Right: The $J-K_s$ vs. $K_s$ color magnitude diagram, the grey points are stars with $K_s < $14 mag within 5 arcmin, the blue points are stars within 0.9 arcmin from the cluster center, the red points stand for the star with similar ($\pm 1$ mas$/$yr).
}  
}} 
\label{pm}
    \end{center}
  \end{figure*}

At the distance of the cluster given by \cite{Froebrich2007b}, $d_{\odot} = 9.1$\,kpc, the five-year VVV baseline is not enough to provide a precise separation of the cluster members and the field stars by proper motion analysis.
Nevertheless, using a proper motion pipeline developed by \cite{Smith2015}, we were able to estimate the mean relative motion of the cluster. Briefly, the proper motion pipeline splits each array of the VVV catalog at every epoch into 5$\times$5 sub-arrays. The coordinate system of each sub-array at each epoch is transformed to the master epoch (the epoch with best seeing) coordinate system using a linear fit and an iterative rejection of astrometric reference sources with significant ($>3 \sigma$) residuals. The proper motions are fit in array coordinate space using the Levenberg-Marquardt algorithm through the \textsc{curve\_fit} function of the \textsc{scipy.optimize} package. Individual epochs are iteratively removed if they are ($>3 \sigma$) from the fit. The final array coordinate motions are converted to equatorial coordinates using the astrometric fit parameters of the master epoch, which are produced by CASU and contained in the FITS header information. The mean uncertainty of the method up to $K_s=$14 mag is under 1 mas/yr on average. Thus, we have selected only stars with $K_s < $14 mag and radius 0.9 arcmin from the cluster center determined in the previous paragraph (red points in Figure~\ref{pm}, left) and compare them with the sources within 5 arcmin in the same magnitude interval (grey points) on the $\mu_{\alpha \cos\delta}$ vs. $\mu_{\delta}$ proper motion diagram. As can be seen, the cluster members are mostly concentrated in a clump in this proper motion diagram, while field stars are much more spread out. The values resulting from the Gaussian fits of the proper motion distributions are $\mu_{\alpha \cos\delta}$= $-$1.3$\pm$0.2 and $\mu_{\delta}$= $+$1.8$\pm$0.3 for the mean relative proper motion of the cluster (expressed in mas/yr; the errors correspond to the fitting errors). 
By taking all stars within a $\pm 1$ mas$/$yr box centered on the proper motion clump and highlighting them in a $J- K_s$ vs. $K_s$ color magnitude (Figure~\ref{pm}, right), we show that the method predominantly select the giants that are cluster members and excluded field dwarfs.

\subsection{Memberships of the spectroscopic targets}\label{mem}

\begin{figure}
     \begin{center}
			\hspace{5cm}
      \includegraphics[scale=0.37]{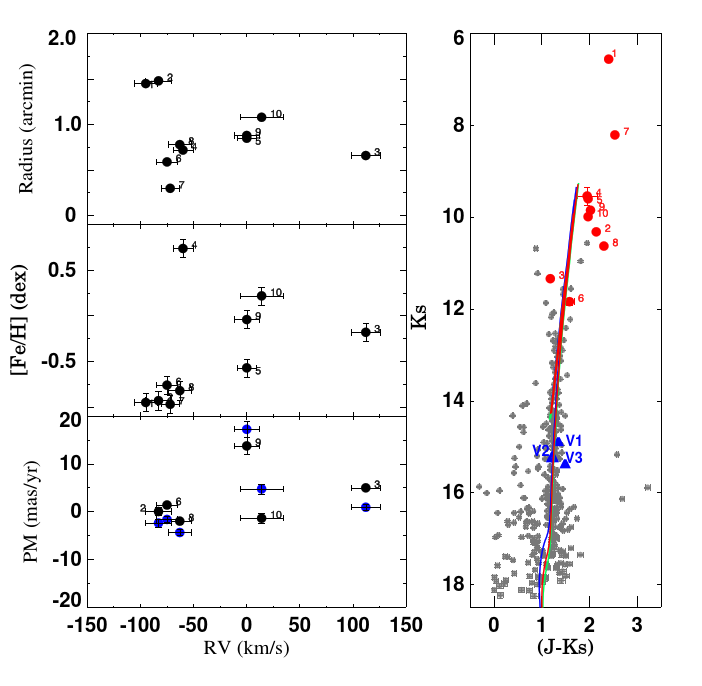}
      \caption[GC3]{ Left: Distance of the spectroscopic target stars with respect to the cluster center in arcmin, the measured [Fe/H] in dex, the relative proper motion $\mu_{\alpha \cos\delta}$ (black) and $\mu_{\delta}$ (blue) in mas/year as a function of the RV in km/s. The error bars represent the random errors of the measurements. 
Right: The ($J-K_s)$ vs. $K_s$ CMD of the statistically decontaminated most probable cluster members. The spectroscopically observed stars are plotted as red circles while RR\,Lyrae stars are plotted as blue triangles. The error bars represent the photometric errors and the isochrones with ages $t = 9.8, 10$ and 12.6\,Gyr are overplotted as a blue, red and green solid lines, respectively.}
\label{fig_feh_vr}
     \end{center}
   \end{figure}

The spectroscopic targets were selected from statistically decontaminated photometry (see Sect. 2). However, our statistical decontamination methods are not designed to definitively determine cluster membership. As a result, we inevitably obtained some spectra of field stars. Cluster membership could be determined more accurately from the proper motion, radial velocity and metallicity histograms. The left panel of \figref{fig_feh_vr} shows the distance of the star from the cluster center, metallicity and relative proper motion vs. measured radial velocities of the spectroscopically observed stars. It was not possible to use the relative proper motion  of spectroscopically observed stars to separate field stars from cluster  members, because only six of them have measured proper motions, the rest are saturated in the VVV images. In addition, using the \cite{Bonatto2011} algorithm, we fitted the radial velocity and [Fe/H] distributions with a Gaussian profile. The resulting values of the Gaussian fits are as [Fe/H]=-0.98$\pm 0.05$ (dex), $\mu$=1.4$\pm 0.4$\,mas/yr and RV=-69$\pm 3$\,km/s.

The analysis of such constructed diagrams shows that stars number 3, 4, 5, 9 and 10 are definitely outliers from the fitted distributions and thus are considered field stars. The stars number 1, 2, 6, 7 and 8 are probable cluster members. Looking at the position of the stars numbers 1 and 7 on the CMD (\figref{fig_feh_vr}, right panel) we can speculate that most probably these stars are red variables, but they are saturated in VVV images so it is not possible to test this suggestion. 

Considering only cluster member stars (number 1, 2, 6, 7 and 8), the mean metallicity and radial velocity for GC\,03 are calculated as $<$[Fe/H]$> = -0.9 \pm 0.2$ and $<v_{\rm r}> = -78 \pm 12$\,km/s. Such calculated [Fe/H] value is in agreement with the metallicity interval reported in \citet{Froebrich2007a} and lower than \cite{Kharchenko2013} estimate. However, we have to point out that this is a first spectroscopic determination, while the above cited authors used photometric calibrations to derive the metallicity.

\subsection{Variable stars}

\begin{figure}
     \begin{center}
     \hskip 0.5cm
     \includegraphics[scale=0.5]{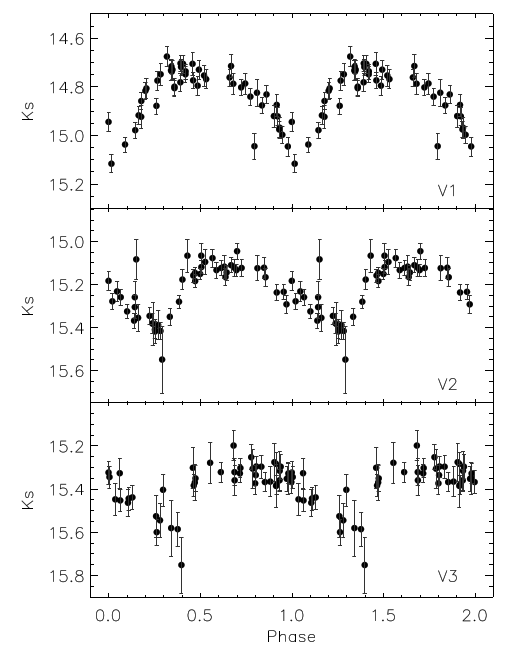} 
      \caption[GC3]{$K_s$ light curves of the RR\,Lyrae stars found in VVV likely associated with GC\,03.}
\label{var}
     \end{center}
   \end{figure}

Using the VVV five year database we searched for variability within 15\,arcmin radius from the cluster center. Forty nine $K_s$ epochs are available between 2010 and 2014, the searching method was focused on variations in the light curves greater than 0.2 in $K_s$ (conservative upper limit of our photometry, including the photometric errors, calibration to the standard system, etc.). Three variables are found, namely V1, V2 and V3, with mean $K_s$ magnitudes $<K_s> = 14.83, 15.22$ and 15.38 mag, respectively. Using the information potential metric developed in \cite{Huijse2011} we found their periods in the range $ 0.48 < P < 0.55$\,days. Thus, they are classified as RRab Lyrae stars. Their coordinates are listed in \tabref{var} and their position in the CMD is shown in \figref{fig_feh_vr}. Note, that we plotted the mean $K_s$ magnitude and random phase of corresponding J ones. The light curves are shown in \figref{var}.

Assuming that these RR~Lyrae stars are associated with GC\,03, the period--luminosity calibration from \citet{Muraveva2015} gives a mean heliocentric distance to the cluster of $d_{\odot} = 10.8 \pm 0.4 $\,kpc. The RR Lyrae calculated distance value is bigger than previously found by Froebrich et al. and Kharchenko et al. 2013 and we will adopt $d_{\odot} = 10.8 \pm 0.9 $\,kpc,  as our final distance determination, adding quadratically the conservative limit of the errors from photometry, calibration, etc.

\subsection{Isochrone fitting}

To constrain the age, we generated a grid of the theoretical isochrones from Padova library \citep{Bressan2012} with ages between 6--12\,Gyr with metallicity fixed on the value derived above.  We also fixed the distance to the above calculated from RR Lyrae stars value. Then, we fitted these isochrones to the observed CMD and reasonable matches were obtained within range of ages.  Given that our photometry does not reach the turn-off level of the population, the isochrones with ages in the range 10--12.5\,Gyr are suitable to reproduce the observed CMD, which is consistent with the presence of RR\,Lyrae stars. Therefore, we adopt a mean value of $t = 11.3\pm$ 1.2\,Gyr with a distance modulus of  $(m-M)_0 = 15.17\pm0.5$ and $A_{\rm K} = 0.37$. The isochrones are overplotted in \figref{fig_feh_vr} and the parameters derived are shown in \tabref{table_results}.
 
Based in our heliocentric distance, we estimate that the distance with respect to the plane for GC\,03 is $|z| \sim 350$\,pc. Given that the scale heights for the thin and thick disks are set, respectively, in $h_{\rm z} = 300$ and 900\,pc \citep{Juric2008}, we confirm that this cluster is immersed in the Galactic disk. Its location in the Galaxy together with the parameters derived above might help us establish the nature of GC\,03. While young stellar clusters ($t < 500$\,Myr) are concentrated towards the plane, older open clusters (OCs) show an exponential distribution with scale heights greater than $375$\,pc and as a function of age \citep{Janes1994,Joshi2005,Bonatto2006,Buckner2014}. In this context, GC\,03 might belong to the Galactic OC population becoming one of its oldest members known, $\sim 2$\,Gyr older than NGC\,6791 \citep{Brogaard2012} and with a similar age to that of Berkeley\,17 \citep{Krusberg2006}. On the other hand, the mean [Fe/H] derived for old ($t > 1$\,Gyr) OCs in the Galaxy contained in the WEBDA\footnote{http://webda.physics.muni.cz} compilation is [Fe/H] = $-0.2 \pm 0.3$, where the old Berkeley\,17 presents a metallicity of [Fe/H] = $-0.3$. GC\,03 is thus metal-poorer than the bulk of OCs, including coeval clusters.

The radial distribution of stellar clusters in the Milky Way might also provide clues about the classification of GC\,03.  With the distance derived via isochrone fitting and using the transformation matrix shown in Appendix\,A in \cite{Pasetto2011}, we estimate that GC\,03 is located at $d_{\rm G} = 4.2$\,kpc from the Galaxy center. OCs older than 6\,Gyr are scarse in the Mily Way and all the clusters older than 1\,Gyr are located at $d_{\rm G} > 6$\,kpc \citep{Bonatto2006}, although a fraction of clusters might remain undetected in high-extinction areas. Diverse theoretical studies have established that the dissolution time ($t_{\rm dis}$) for clusters with $M < 10^{4}$\,M$_{\odot}$ is lower than a few hundreds of Myr \citep{Lamers2005}, while for OCs with $M \sim 10^{4}$\,M$_{\odot}$ it has been estimated a $t_{\rm dis} \sim 2$\,Gyr \citep{Bergond2001,Gieles2006}. Therefore, it is expected that all the OCs observed nowadays in the inner kpc of the Galaxy are young and massive. Indeed, massive OCs with ages of a few Myr have been found with $R_{\rm GC} < 4$\,kpc \citep[e.g.][]{Figer2002,Crowther2006}. In this context, GC\,03 could only survive as an OC at $R_{\rm GC} < 4$\,kpc if it was as massive as a Galactic GC. 

In the case of the Galactic GC population, a significant fraction of the systems lie in the volume defined by $R_{\rm GC} = 4$\,kpc. From the 157 clusters contained in the \cite{Harris2010} catalog, 64 GCs (40$\%$) have $R_{\rm GC} < 4$\,kpc and half of them are found at low Galactic latitudes ($|b| < 5^{\circ}$). That latter group of GCs has a mean metallicity of [Fe/H] $-0.8 \pm 0.4$, which is a similar value to that derived from our spectra for GC\,03. Therefore, the position of GC\,03 in the Galaxy and its metallicity, similar to that of the bulk of bulge GCs allow us to confirm the globular nature of that cluster.

GC\,03 might belong, following the scheme suggested in the classical works by \cite{Zinn1993} and \cite{Armandroff1989},  to the population of disk GCs, rather than to the halo/accreted globulars found in the outer regions of the Galaxy. A prototypical globular included in that metal-rich subgroup and associated with the Milky Way thick disk is 47\,Tuc (NGC\,104). That GC has a metallicity of [Fe/H] = $-0.7$ with an age of 9.9\,Gyr \citep{Carretta2009,Hansen2013}, which are values similar to those derived in this work. This suggests that GC\,03 might have been formed together with the disk stellar populations and the metal-rich component of Galactic GCs. 

\subsection{Radial density profile and spatial distribution}

\begin{figure}
     \begin{center}
      \includegraphics[scale=0.75]{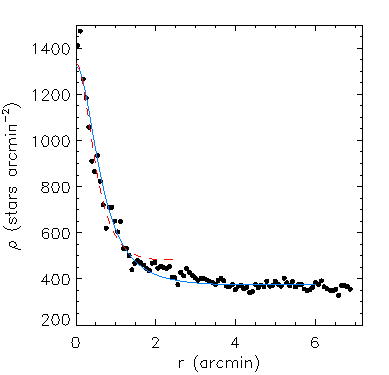}
      \caption[Radialprofile]{Radial density profile derived for GC03. The red dashed line indicates the best King profile fitted using only those stars with $r<1\farcm7$ while the solid blue line corresponds to the fitting including those stars with $r<7$'. } 
\label{radialprofile}
     \end{center}
   \end{figure}

We studied the cluster radial density profile, calculating the stellar density in concentric annuli of width $4\arcsec$, centered on the cluster coordinates derived above and using VVV photometry. The results are shown in \figref{radialprofile}. If the analysis is limited to stars with $K_s<14.7$ and distances from the center $r<1\farcm7$, as done by \cite{Froebrich2007b}, the radial decay of the stellar density is well represented by a King profile, with $r_c=0\farcm6\pm0.2$ and $r_t=2\farcm2\pm0.7$ as the best-fit core and tidal radii, respectively. These values are slightly larger than those found by Froebrich et al., but comparable within errors, indicating that our VVV stellar counts are consistent with theirs. Nevertheless, when the profile is extended to $r=7\arcmin$, it becomes evident that the background is overestimated by this fit, because the stellar density keeps decreasing beyond $r=1\farcm7$. The profile is very flat for $r>4\arcmin$, suggesting that small-scale fluctuations are negligible. However, the tidal radius diverges to huge values if the fit is extended to $r>2\arcmin$. Identical results are found extending the stellar counts down to $K_s<16.5$. The cause of this unrealistic result is that the density in the range 2--3$\arcmin$ is too high than expected for a King profile. The tidal radius can be reduced to 9$\arcmin$ excluding the region 2--2.5$\arcmin$ from the fit and a minimum value of $4\farcm8$ is obtained with the exclusion of the 2--3.5$\arcmin$ range. The core radius barely changed in any of these fits and it was always found in the range $r_c=0.75\pm0.05\arcmin$. Our results are consistent with those by \cite{Kharchenko2013}  where it was also found a very large tidal radius ($8\pm2\arcmin$), although they do not provide the details of their fit.

\begin{figure}
     \begin{center}
      \includegraphics[scale=0.5]{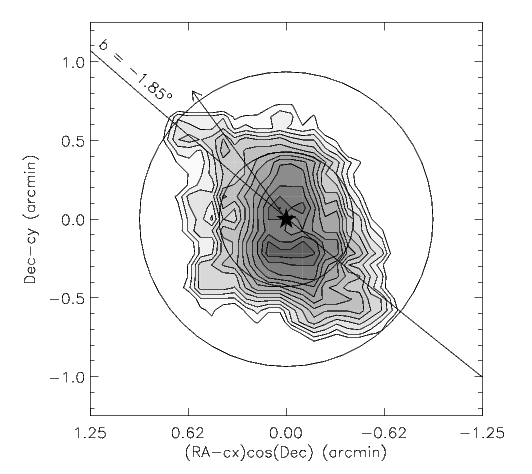}
 \hspace{-0.5cm}
      \caption[Map]{Stellar density contours showing the distribution in the sky of stars belonging to GC\,03 according to the decontaminated CMD. The inner and outer circles correspond to the distances $r_{\rm 1} = 0.45$\,arcmin and $r_{\rm 2} = 0.95$\,arcmin respectively. The constant Galactic latitude level at $b = -1.85^{\circ}$ is overplotted with a solid line while the direction towards the Milky Way center is indicated with an arrow.} 
\label{densitymap}
     \end{center}
   \end{figure}

Our analysis shows that GC03 is likely more extended than the estimates of \cite{Froebrich2007b} reported in the \cite{Harris2010} catalog. However, it is also clear that the cluster density profile is not well represented by a King function. This behavior is typical of a cluster suffering intense tidal stress \citep[see, e.g.,][]{MoniBidin2011,Carballo-Bello2012}. Indeed, \figref{densitymap} shows the  density map generated with stars potentially associated with the GC, with a nearly symmetrical and elongated distribution of stars similar to that observed in GCs with known tidal tails \citep[e.g. Pal\,14; see][]{Sollima2011}. 

The region with higher density in \figref{densitymap} seems to be contained in an area with radius $r_{\rm 1}= 0\farcm45$ (inner circle in that plot), while the density of stars beyond $r_{\rm 2} = 0\farcm95$ (outer circle) is negligible. The distance $r_{\rm 1}$ is consistent with the core radius derived above for this cluster when only stars with $r < 1\farcm7$ are considered, while the obtained King tidal radius is much larger than $r_{\rm 2}$. At the distance of GC\,03, the radius of its observed core is $r_{\rm 1} = 1.5$\,pc and the tentative inner tidal tails expand up to $r_{\rm 2} = 3.2$\,pc from the cluster center. 

\begin{table}
 \centering
  \caption{Parameteres derived for GC\,03. Note that proper motions are relative motions (see section 3.2).}
  \begin{tabular}{lr}
  \hline
  Parameter \ \ \ & \ \ \ Result\\
  \hline
\\
 $[Fe/H]$  & $-0.9 \pm 0.2$\\
 $d_{\odot}$ [kpc] & $10.8 \pm 0.9$\\
 $d_{G}$ [kpc] & $4.2 \pm 0.1$ \\
 $t$ [Gyr] & $11.3 \pm 1.2$\\
 $v_{\rm r}$ [km\,s$^{-1}$] & $-78 \pm 12$\\
 $\mu_{\alpha \cos\delta}$ (mas\,yr$^{-1}$) & $-1.3 \pm 0.2$ \\
 $\mu_{\delta}$ (mas\,yr$^{-1}$) & $1.8 \pm 0.3$\\
  $r_{\rm c}$ (arcmin) & 0.75 $\pm$ 0.05 \\
 $r_{\rm t}$ (arcmin) & $>$4.8 \\

\end{tabular}
\label{table_results}
\end{table}

\section{Conclusions}

We have derived fundamental parameters for the stellar cluster 2MASS-GC\,03 using MMIRS photometry and spectroscopy and VVV data.  Out the 10 stars for which we have obtained spectra, 5 were considered as cluster members based on their proper motions derived from VVV photometry. We have estimated metallicities and radial velocities for field and candidate stars and set mean values for the cluster in $<[Fe/H]> = -0.9$ and $<v_{\rm r}> \sim -78$\,km\,s$^{-1}$. With that information, we have established that 2MASS-GC\,03  is an intermediate-age system with $t = 11.3$\,Gyr at $d_{\odot} = 10.8$\,kpc. Its location within the Galaxy and overall properties (summarized in Table~2) confirms its globular nature.

\section*{Acknowledgements}
This paper includes data gathered with the 6.5 meter Magellan Telescopes located at Las Campanas Observatory, Chile (program ID: CN2014A-84). JAC-B received support from CONICYT Gemini grant from the Programa de Astronom\'ia del DRI Folio\,32130012 and CONICYT-Chile grants FONDECYT Postdoctoral Fellowship 3160502. SRA and CMB were supported by the Fondecyt project number 3140605 and 1150060, respectively. JA-G acknowledges support from the FIC-R Fund, allocated to project 30321072, and from Fondecyt Iniciaci\'on 11150916. N.M is suported  by CONICYT REDES No. 140042 project. The VVV Survey is supported by ESO, by BASAL Center for Astrophysics and Associated Technologies PFB-06, by FONDAP Center for Astrophysics 15010003, and by the Ministry for the Economy, Development, and Tourism's Programa Inicativa Cient\'{i}fica Milenio through grant IC120009, awarded to The Millennium Institute of Astrophysics (MAS). Partially based on data products from observations made with ESO Telescopes at the La Silla or Paranal Observatories under ESO programme ID 179.B-2002. This publication makes use of data products from the Two Micron All Sky Survey, which is a joint project of the University of Massachusetts and the Infrared Processing and Analysis Center/California Institute of Technology, funded by the National Aeronautics and Space Administration and the National Science Foundation. Thanks to K. Pe\~na-Ram\'irez and J. M. Corral-Santana for their help during the spectra reduction.

\def\jnl@style{\it}                       
\def\mnref@jnl#1{{\jnl@style#1}}          
\def\aj{\mnref@jnl{AJ}}                   
\def\apj{\mnref@jnl{ApJ}}                 
\def\apjl{\mnref@jnl{ApJL}}               
\def\aap{\mnref@jnl{A\&A}}                
\def\mnras{\mnref@jnl{MNRAS}}             
\def\nat{\mnref@jnl{Nat.}}                
\def\iaucirc{\mnref@jnl{IAU~Circ.}}       
\def\atel{\mnref@jnl{ATel}}               
\def\iausymp{\mnref@jnl{IAU~Symp.}}       
\def\pasp{\mnref@jnl{PASP}}               
\def\araa{\mnref@jnl{ARA\&A}}             
\def\apjs{\mnref@jnl{ApJS}}               
\def\aapr{\mnref@jnl{A\&A Rev.}}          

\bibliographystyle{mn2e}
\bibliography{draft}

\begin{table*}
 \centering
  \caption{Parameters derived for the RR\,Lyrae found in GC\,03.}
  \begin{tabular}{lccclll}
  \hline
  ID & RA [J2000] & Dec [J2000] & $<Ks>$& Ampl.& P (d) & $d_{\odot}$ (kpc) \\
  \hline
  V1 & 253.0405607 & -47.077880 & 14.88&0.45 &0.84310 & 11.2 \\
  V2 & 253.0267025 & -47.717799 & 15.21&0.50 &0.54171 & 10.5 \\
  V3 & 252.8666458 & -46.995242 & 15.36&0.55 &0.55330 & 10.8 \\
\end{tabular}
\label{rrlyrae}
\end{table*}

\end{document}